\documentclass{article}
\usepackage[utf8]{inputenc}
\usepackage{makeidx}  
\usepackage{graphicx}
\usepackage{balance}
\usepackage{listings}
\usepackage{multirow}
\usepackage{threeparttable}
\usepackage{booktabs}
\usepackage{epstopdf}
\usepackage{amsmath,amsfonts,amssymb,mathrsfs,amsthm}
\usepackage{float}                              
\usepackage{multirow}
\usepackage{caption}
\usepackage{subfig}
\usepackage{footnote}
\usepackage{color}
\usepackage{pifont}
\usepackage{acronym}
\usepackage{pgfplots}
\usepackage{eurosym}
\pgfplotsset{compat=newest}
\usepackage{textcomp}
\usepackage{pdflscape}

\acrodef{PV}{photovoltaic}
\acrodef{HEMS}{home energy management system}
\acrodef{NILM}{non-intrusive load monitoring}
\acrodef{HMM}{hidden Markov model}
\acrodef{FHMM}{fractional hidden Markov model}
\acrodef{PF}{particle filtering}
\acrodef{ILM}{intrusive load monitoring}
\acrodef{FSM}{finite state machine}
\acrodef{RMSE}{root mean square error}
\acrodef{BLH}{battery-based load hiding}
\acrodef{LLH}{load-based load hiding}
\acrodef{BE}{best-effort}
\acrodef{NILL}{non-intrusive load leveling}
\acrodef{SF}{stepping framework}
\acrodef{SOC}{state of charge}
\acrodef{H}{hamming loss}
\acrodef{ACC}{accuracy}
\acrodef{GREEND}{Green Electric Energy Dataset}

\definecolor{orange}{rgb}{1,0.5,0}

\makeatletter
\def\blfootnote{
 \xdef\@thefnmark{}\@footnotetext
 }
\makeatother

\begin{document}
	\setlength{\baselineskip}{.965\baselineskip}

\title{Load Hiding of Household's Power Demand}

 \author{
Dominik~Egarter{\small $~^{1}$}, Christoph~Prokop{\small $~^{2}$}, and~Wilfried~Elmenreich{\small $~^{1}$}\\
Institute of Networked and Embedded Systems\\
 Alpen-Adria-Universit\"at Klagenfurt, Austria\\
 {\small $~^{1}$}\{\emph{name.surname}\}@aau.at\\
 {\small $~^{2}$}christoph.prokop@student.tugraz.at
 }

\maketitle              
\blfootnote{ 
This work was performed in the research cluster Lakeside Labs funded by the European Regional Development Fund, the Carinthian Economic Promotion Fund (KWF), and the state of Austria under grants 20214/22935/34445 (Smart Microgrid Lab).
}
\begin{abstract}
With the development and introduction of smart metering, the energy information for costumers will change from infrequent manual meter readings to fine-grained energy consumption data. 
On the one hand these fine-grained measurements will lead to an improvement in costumers' energy habits, but on the other hand the fined-grained data produces information about a household and also households' inhabitants, which are the basis for many future privacy issues.
To ensure household privacy and smart meter information owned by the household inhabitants,  load hiding techniques were introduced to obfuscate the load demand visible at the household energy meter.
In this work, a state-of-the-art \ac{BLH} technique, which uses a controllable battery to disguise the power consumption and a novel load hiding technique called \ac{LLH} are presented. 
An \ac{LLH} system uses an controllable household appliance to obfuscate the household's power demand.
We evaluate and compare both load hiding techniques on real household data and show that both techniques can strengthen household privacy but only \ac{LLH} can increase appliance level privacy.
\end{abstract}
%
%
%
\section{Introduction}

In the context of smart grids and smart metering, the term privacy is becoming of high interest and is much discussed. 
Smart meters are accurate monitoring units providing fine-grained demand measurements in which these monitoring results disclose user behavior which could be extracted by smart algorithms and techniques.
The foundation of algorithms to extract energy consumption information was set in 1992 with the introduction of \ac{NILM} \cite{Hart1992}.
\ac{NILM} is a single-point metering approach which detects and identifies appliances in the total power demand of households.
It uses appliance specific characteristics and smart classification approaches to  identify appliances and sense at which point in time which appliances were running.
State-of-the-art approaches \cite{Zeifman2011, Egarter2013BuildSys} depend on the granularity of the measurements.
With $1s$ measurement granularity \ac{NILM} approaches can disaggregate around $10$ different appliances \cite{CarrieArmel2013}.
With information of the power demand habits on appliance level, it is possible to extract user behaviors and habits by activity recognition and user profiling \cite{Nguyen2013244,Lisovich2010}.
An extreme example for analysing the energy data on appliance level is shown in \cite{Greveler2012}. 
In this work a smart meter is used to identify the multimedia content of a TV.
Potentially interested stakeholders are presented in \cite{Skopik} such as the energy utility, creditors, press and marketing/advertisements partners, in an extreme case even criminals.

The loss of privacy by load disaggregation and energy mining is a huge upcoming smart grid and society issue which enforces the need of privacy preserving techniques, which can be divided into the following three possibilities \cite{Skopik}:
\begin{enumerate}
  \item \textit{Anonymization of metering data}: The metering data and costumer identity are separated by a third-party id \cite{Efthymiou2010}.
  \item \textit{Privacy-preserving metering data aggregation}: Metering data is geographically encapsulated by aggregating the metering data of co-located consumers \cite{Li2011}
   \item \textit{Masking and obfuscation of metering data}: Masking the power demand by adding or withdrawing the to the meter visible energy demand with the help of rechargeable batteries \cite{Yang2012} or controllable loads.
\end{enumerate}
To ensure household privacy without any interaction by third-parties or neighbours and to keep information at the owner side, this paper concentrates on obfuscating metering data. 
Obfuscating the metering data is usually done by controllable batteries and is called \ac{BLH}.
A \ac{BLH} system charges and discharges the battery at strategic times to flatten the household's energy demand.
In this work we introduce a novel obfuscation approach of power draws called \acf{LLH}.
It uses energy-intensive household loads which are controllable, have a daily power consumption and are not user driven.
A common household device which meets these requirements is an electric water boiler. 
In this paper, we describe controlling a boiler by randomly turning it on and off  with the constraint to meet a given daily power consumption.
\ac{LLH} is obfuscating the power demand by putting noise to the power draw in contrast to \ac{BLH}, which is trying to flatten the power draw.\newline
The aim of this work is to test the novel introduced \ac{LLH} technique to a state-of-the-art \ac{BLH} technique.
The obfuscation performance for both load hiding techniques is tested by a state-of-the-art \ac{NILM} algorithm and by the evaluated error between the real and the obfuscated power draw.
The tests are done on real household consumption data using a realistic model of  battery and electric boiler. \newline
The remainder of this paper is organized as follows: Section \ref{sec:BLH} and \ref{sec:LLH} are presenting basics about \ac{BLH} and \ac{LLH} whereas Section \ref{sec:settings} describes the evaluation settings of the experiments such as the used dataset and evaluation metrics and the configuration and implementation of the \ac{BLH} and \ac{LLH} system. 
In Section \ref{sec:results}, the results of the simulations for \ac{BLH} and \ac{LLH} are shown and are evaluated by the achieved obfuscation difference and the ability to detect appliances with a proposed \ac{NILM} technique in the obfuscated power draw.
Finally, the results and the pros and cons for both load hiding techniques are discussed in Section \ref{sec:discussion} and concluded in Section \ref{sec:conclusion}.

\section{Battery-Based Load Hiding}\label{sec:BLH}
\begin{figure}[h!]
	\centering
	\includegraphics[width=80mm]{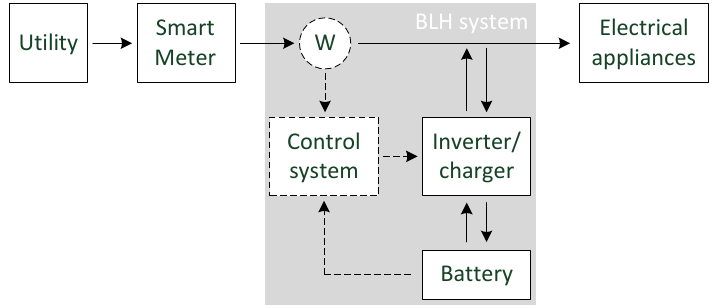}
	\caption{Schematic representation of a \ac{BLH} system}
	\label{blh_system}
\end{figure}

The first proposal to mask power demand using a rechargeable battery was presented in \cite{Kalogridis2010}. The idea bases on the installation of an intelligent \ac{BLH} system between the smart meter and the internal wiring such as it is plotted in Figure~\ref{blh_system}.

\begin{figure}[h!]
	\centering
	\includegraphics[width=85mm]{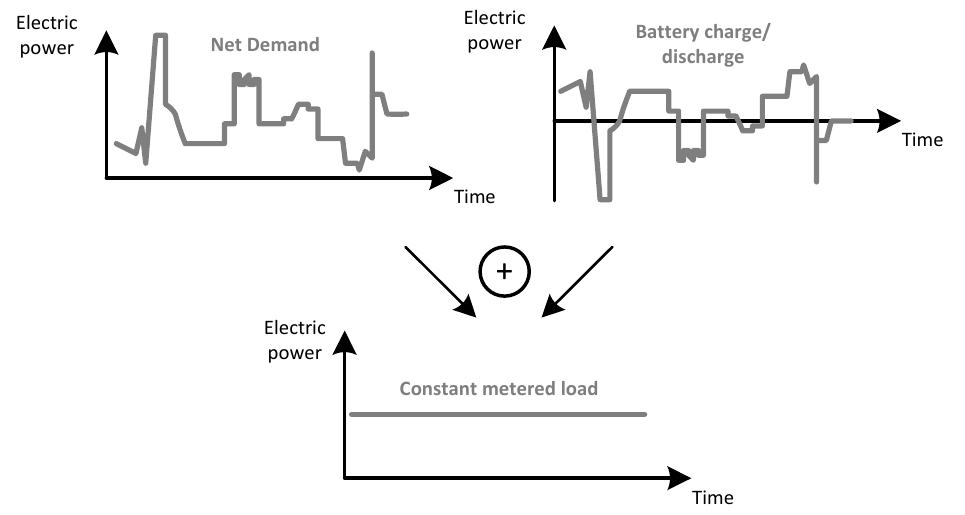}
	\caption{\ac{BLH} approach to flatten the metered load \cite{McLaughlin2011}}
	\label{blh_method}
\end{figure}

The \ac{BLH} system charges or rather discharges the battery at strategic times to modify the metered load, i.e. the electric active power demand that is observed by the smart meter. The aim is to hide or obscure load signatures, so that appliance usage events and usage patterns cannot be detected by \ac{NILM} \cite{Kalogridis2010}. The first proposals of \ac{BLH}-algorithms try to maintain a constant metered load. Any changes in net demand, which is the household's active power demand except the \ac{BLH} system, should be covered by the battery to flatten the energy consumption observed by the smart meter. Such an approach is diagrammed in Figure~\ref{blh_method}.

Unfortunately, in practise there are physical limitations of batteries like a maximum charging and discharging rate or the limited capacity of the battery. Taking these battery constraints and battery prices into account, the installation of a \ac{BLH}-system that is capable to hide all usage events under all circumstances would be very costly. This leads to an optimization problem minimizing the leakage of information using feasible battery sizes. Several algorithms have been proposed to maintain the metered load that is transmitted to the utility constant as long as possible \cite{Yang2012}, such as the \ac{BE} algorithm, the \ac{NILL} algorithm and the \ac{SF}. According to \cite{Yang2012} the most promising algorithm is the \ac{SF} using \textit{Lazy Stepping~2} algorithm, which will be used as the representative \ac{BLH} algorithm for further evaluations.

The \ac{SF} \cite{Yang2012} makes the metered load to be integer multiples of a constant value, which is the minimum of either the maximum charge and discharge rate. For any possible net demand there exists a multiple of this constant satisfying the battery constraints. For each level of net demand one can choose either the level just higher than net demand, which will charge the battery, or the level just lower than net demand which will discharge the battery. If the battery's \ac{SOC} gets too high, then the system chooses the lower level and the reverse happens when the battery's state gets too low. During normal operation the decision of whether to choose the upper or lower level is task of the \ac{SF}. The authors suggest four algorithms: \textit{Lazy Stepping 1}, \textit{Lazy Stepping 2}, \textit{Lazy Charging} and \textit{Random Charging}. With regards to the authors the most successful approach is \textit{Lazy Stepping 2}. It keeps the metered load constant if possible, otherwise it randomly chooses the upper or rather the lower level.


\section{Load Based Load Hiding}\label{sec:LLH}
\begin{figure}[h!]
	\centering
	\includegraphics[width=80mm]{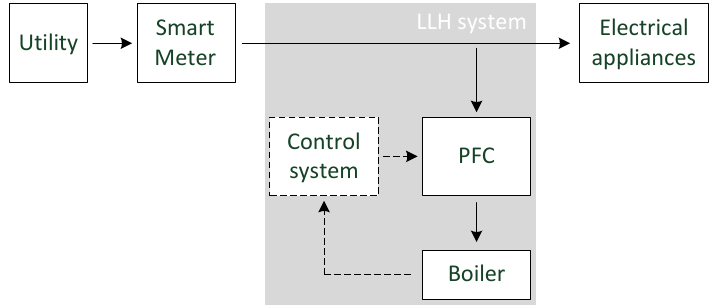}
	\caption{Schematic representation of a \ac{LLH} system}
	\label{llh_system}
\end{figure}

\begin{figure}[h!]
	\centering
	\includegraphics[width=85mm]{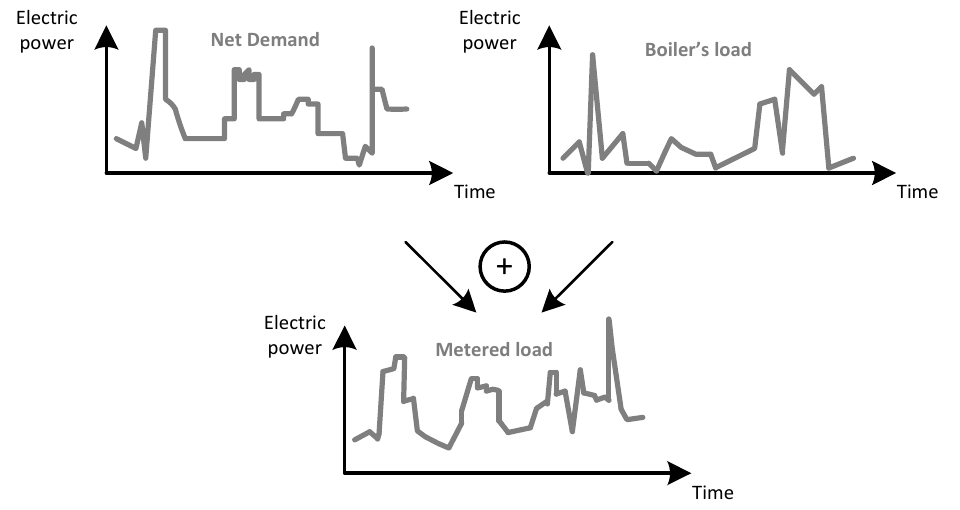}
	\caption{\ac{LLH} approach to add noise}
	\label{llh_method}
\end{figure}

Similar to a \ac{BLH} system, one could realize an obfuscating system by using a variable load instead of the battery system. A variable load should be a powerful interruptible process that is not time-critical, not directly user driven and adjustable in its power consumption. Such a process could be a domestic electric hot water boiler, an electric heater, or perhaps even an electric vehicle charger.
The device is assigned by a daily target energy consumption, i.e., an amount of energy the device is supposed to spend during a day in order to fulfill its target function.
The device is not bound to certain times when to use energy but is limited by a maximum power.
 In the remainder, systems using such a variable load will be shortened by \acf{LLH}.
A \ac{LLH} system has the major aim to increase (but not decrease as for \ac{BLH}) the metered load level compared to the corresponding net demand. This is limited by both, the maximum power of the appliance used as a variable load and by the necessary energy consumption of the device during a day.
A novel implementation of a \ac{LLH} system does not require any changes in the households internal wiring and furthermore necessitates no extra measurements like the actual level of net demand (demand apart from the variable load) which is necessary in case of a BLH system. The target model in this work is a completely passive electric boiler without any knowledge of the internal wiring, the appliances in use, net demand or the metered load. Figure~\ref{llh_system} plots the schematic of such a system. The control system adjusts the current power consumption via a phase-fired controller (PFC) that modifies the voltage level depending on the customer's needs and the set temperature of the boiler. Note that $P \sim U^2$ as $P=\frac{U^2}{R}$ where $R$ can be assumed to be constant.
Without any data of net demand, maintaining a constant metered load is impossible. Additionally, holding a constant value like under \ac{BLH} would necessitate some kind of forecast to fill the gap between net demand and the constant load to still meet the targeted energy level at the end of each day without leaking too much information. 
The basic idea of this proposal is to overlay net demand by a probabilistic signal, i.e.\ artificial created noise, which impedes the detection of the appliance’s states. Figure~\ref{llh_method} plots net demand that is overlayed by a probabilistic load of the electric boiler.
The basis of this artificial noise is a probability distribution function, such as a beta distribution. The realizations must lie within the interval $[0 \: P_{max}]$ where $P_{max}$ is the maximum power of the variable load. 
With regards to the preprocessing of the \ac{NILM} algorithm, that applies filters such as a running median filter, a higher level of randomization of the noise should decrease the efficacy of the \ac{NILM} algorithm. We expect that a modified beta distribution can increase the level of privacy protection. 

\section{Evaluation Settings}\label{sec:settings}
\subsection{Implementation}
\subsubsection{\ac{LLH}-Settings}
The simulation model of the \ac{LLH} system is based on several simplifications.
This paper does not implement a dynamic model of the electric boiler but only the maximum power of the boiler of $1600\:W$ and a daily target energy consumption without considering any further losses or devices. 
For the sake of simplicity, we assume a daily target energy consumption disregarding the temperature, amount or time of use of the hot water as the boiler's end product.
In detail, the daily target energy consumption is set to four scenarios $[2.5, 5, 7.5, 10]\: kWh$.
For simulating \ac{LLH} $P_{max}$ is set to $P_{max}=1600 \: W$. Based upon pre-evaluations, the parameter set of the underlying beta distribution is $\alpha=0.9$, and $\beta$ is derived by the expectation $\mu$: $\beta=\frac{\alpha - \alpha \cdot \mu}{\mu}$. In order to meet the target energy consumption, the expectation of the distribution function and the comparable constant load must be balanced, e.g.: a daily energy target of $5 \: kWh$ can be realized by a constant load of $\mu_{set} = \frac{5000 \: Wh}{24 \: h} = 208.\dot{3} \: W$ but also by realizations of a random variable based on a probability distribution function with the same mean. 
When applying the modified beta distribution, the output varies randomly between $0$ and $P_{max}$. The expectation lies in $[0, P]$ for a random time frame with a maximum of one hour, where $P$ is randomly set between $\frac{P_{max}}{4}$ and $\frac{3 P_{max}}{4}$. If $\mu$ is set higher than $\mu_{set}$ the boiler consumes more energy than it is planned for. Therefore, when setting the following time frame and $\mu$ the energy gap of the realizations compared to the constant load $\mu_{set}$ is analyzed. If the gap exceeds $\pm 0.5 \: kWh$ the new expectation is limited to $[0,\mu_{set})$ or rather $(\mu_{set}, P_{max}]$ depending on whether it was too high or too low. There is a high probability that the daily energy consumption differs from the target consumption. The realized consumption lies in $[4.29,7.8]\: kWh$ for a target energy of $5 \: kWh$ taking the worst case into account.

\subsubsection{BLH-Settings}
\begin{figure}[h!]
	\centering
	\includegraphics[width=80mm]{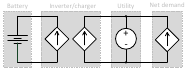}
	\caption{\ac{BLH} simulation model}
	\label{blh_simulation_model}
\end{figure}

The simulation model for \ac{BLH} is modeled in Matlab/Simulink using the SimPowerSystem library that provides a realistic battery model. 
For simplification, this work assumes a purely resistive system considering active power only, which allows the application of a DC model. Furthermore, the inverter/charger combination is idealized by two programmable current sources with zero losses. All other elements are assumed to be ideal as well. The reason for using a simulation model is the application of a realistic battery model that considers the actual \ac{SOC} of the battery. A schematic of the simulation model is plotted in Figure~\ref{blh_simulation_model}.

The two programmable current sources of the inverter/charger combination are controlled by a control system, based on the battery's \ac{SOC}, the battery's voltage level, the actual level of net demand and the \ac{BLH} algorithm's decision. The left source represents the DC-side of the \ac{BLH} system, the right current source represents the utility-side. If the output of the current source on the DC-side is positive, the source acts as a generator and the battery will be charged, if the current is negative
the source acts as a load and the battery gets discharged. Therefore, if the current on the DC-side is positive, the power flows from the utility, that is represented by a voltage source, to the battery. In the following, the utility-side of the inverter/charger combination must act as a load on the secondary branch to consume this energy. Note that the absolute value of the power of both current sources must be balanced. Hence, the absolute values of the current may vary as the battery voltage on the DC-side may vary with the battery's \ac{SOC}. The household's net demand is modeled by another programmable current source.
This work assumes a \ac{BLH} system using a lead-acid storage battery with a nominal voltage of 12\:V. The depth of discharge of the battery is set to 70\%, the \ac{SOC} limits are 20\% or rather 90\%. The initial \ac{SOC} for the first day is assumed to be 55\%, which is the mean of the usable capacity. Both, the maximum charging but also discharging current are set to $0.3\cdot \frac{1}{h}\cdot C$, e.g. $30\:A$ for a $100\:Ah$ type. The rated capacities are set to the simulation cases $[10, 70, 100, 200, 400, 600] \: Ah$.
\subsection{Dataset}
The presented evaluations are based on the dataset \ac{GREEND} \cite{Andrea2014} containing appliance level power measurements of Austrian and Italian households. 
From this dataset we have chosen the household with ID 0 where the residents are a retired couple, spending most of time at home.
We considered seven of the presented household appliances in our evaluations (Table~\ref{tab:BLHResult}) where we have chosen these appliances due to their representative for an household's power demand and their simplicity to be monitored \cite{Carlson2013132}.
Each appliance was monitored separately and was afterwards aggregated to create a realistic household's load profile.
We have chosen a time duration of 14 successive days as an observation window with seconds resolution. 
For preprocessing the time series we used a one-dimensional median filter of order 5.

\subsection{Load Disaggregation Approach}
We used the online load disaggregator based on the work in \cite{Egarter2013BuildSys}, which uses the \ac{PF} approach as a load disaggregator.
The \ac{PF} estimates the appliance state space where each appliance state is represented by the power value consumed by the device.
The decision which appliance is on or off is made by a decision maker based on thresholding.
To represent appliances, their operating states and operating behavior, of each used appliance is modeled by a \ac{HMM}.
The total household's power demand is the aggregated power draw of each appliance for each time instant which is created by a \ac{FHMM}.
The \ac{FHMM} has the advantage to reduce the number of states compared to a simple \ac{HMM} representing the aggregated power draw of each appliance.

\subsection{Metric}

\subsubsection{Load Hiding metric} 
The \ac{RMSE} is well known for measuring the accuracy of forecasting models. In this work, \ac{RMSE} should quantify the deviation between the original time series of net demand and the metered load profile after applying a load hiding system. This should quantify the information loss of the actual level of net demand, as opposed to the load changes in particular. The absolute value of \ac{RMSE} is not significant as it changes with the sample length and other characteristics of the time series considered. In general a higher \ac{RMSE} describes a higher level of privacy protection. The \ac{RMSE} is defined as
$ RMSE = \sqrt{\frac{1}{n} \sum_{i=1}^n (d_i - e_i)^2} $,
were $d_i$ represents the discrete time series of net demand, $e_i$ the discrete time series of the metered load and $n$ the number of time samples.

\subsubsection{Load Disaggregation Metric}
To evaluate a load disaggregator there exists mainly three categories such as the event-based metrics (e.g true positives, true negatives, true positive rate, F-score, etc.), the non event-based metrics (e.g. mean error, \ac{H}, etc.) and the overall metrics (e.g. energy error etc.).
In this work we used the \ac{ACC} to evaluate our dissaggregation results which is described by
$ACC = \frac{TP+TN}{n}$, where $n$ represents the number of time samples, $TP$ (number of times an appliance is correctly detected as on) the number of true positives and $TN$ (number of times an appliance is right detected as OFF) the number of true negatives of the classification process.
The \ac{ACC} is calculated on appliance level and in total by generating the mean of all appliance level \acp{ACC}.
As reference, we calculated the \ac{ACC} for an load disaggregator estimating all appliance states to be off over time. 
A good load disaggregation algorithm is expected to have a significantly better accuracy than this reference.

\section{Results}\label{sec:results}
\subsection{RMSE for \ac{BLH} and \ac{LLH}}

To evaluate the performance of \ac{BLH} and \ac{LLH} in deviating the original time series to the metered time series, the two obfuscation approaches are tested for different simulation cases mentioned.
\begin{figure}[h!]
	\centering
	\includegraphics[width=90mm]{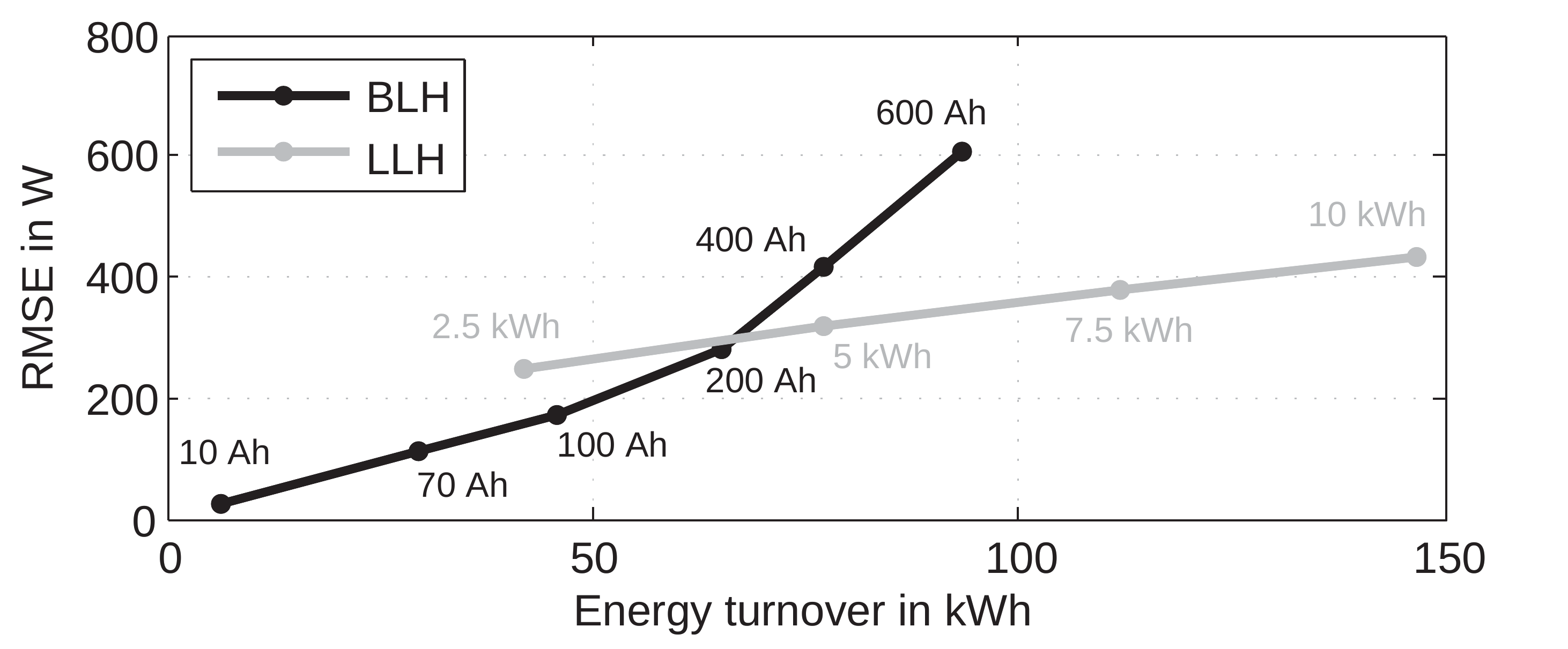}
	\caption{\ac{RMSE} using \ac{BLH} and \ac{LLH} for varying battery capacity and varying daily power consumption}  
	\label{rmse}
\end{figure}
The BLH system varies the rated capacities by $[10, 70, 100, 200, 400,  600]\:Ah$ and the LLH approach varies the daily target energy consumption in between $[ 2.5, 5, 7.5, 10]\:kWh$.
Figure~\ref{rmse} plots the \ac{RMSE} over the energy turnover for both, \ac{BLH} and \ac{LLH}. In case of \ac{LLH} the energy turnover describes the energy consumption over the course of 14 days of the electric water boiler. When applying \ac{BLH} the turnover is the aggregated sum of the absolute value of the energy from and to the battery. The smallest turnover comes along with a $10\:Ah$ battery using \ac{BLH}. When increasing the battery's capacity the \ac{RMSE} increases progressively. When using a variable load the trend is upwards as well, but the \ac{RMSE} increases more linear. Whereas the maximum power of \ac{LLH} is set constant to $1600\:W$, the maximum power of the battery system varies from appr. $36\:W$ for a $10\:Ah$ battery type to appr. $2160\:W$ for the $600\:Ah$ type. For higher capacities this allows \ac{BLH} to distort net demand on a higher level, which in turn yields to a greater \ac{RMSE} compared to \ac{LLH}.

\begin{landscape}
\begin{figure*}
\begin{minipage}{0,32\columnwidth}
		\includegraphics[scale=0.25]{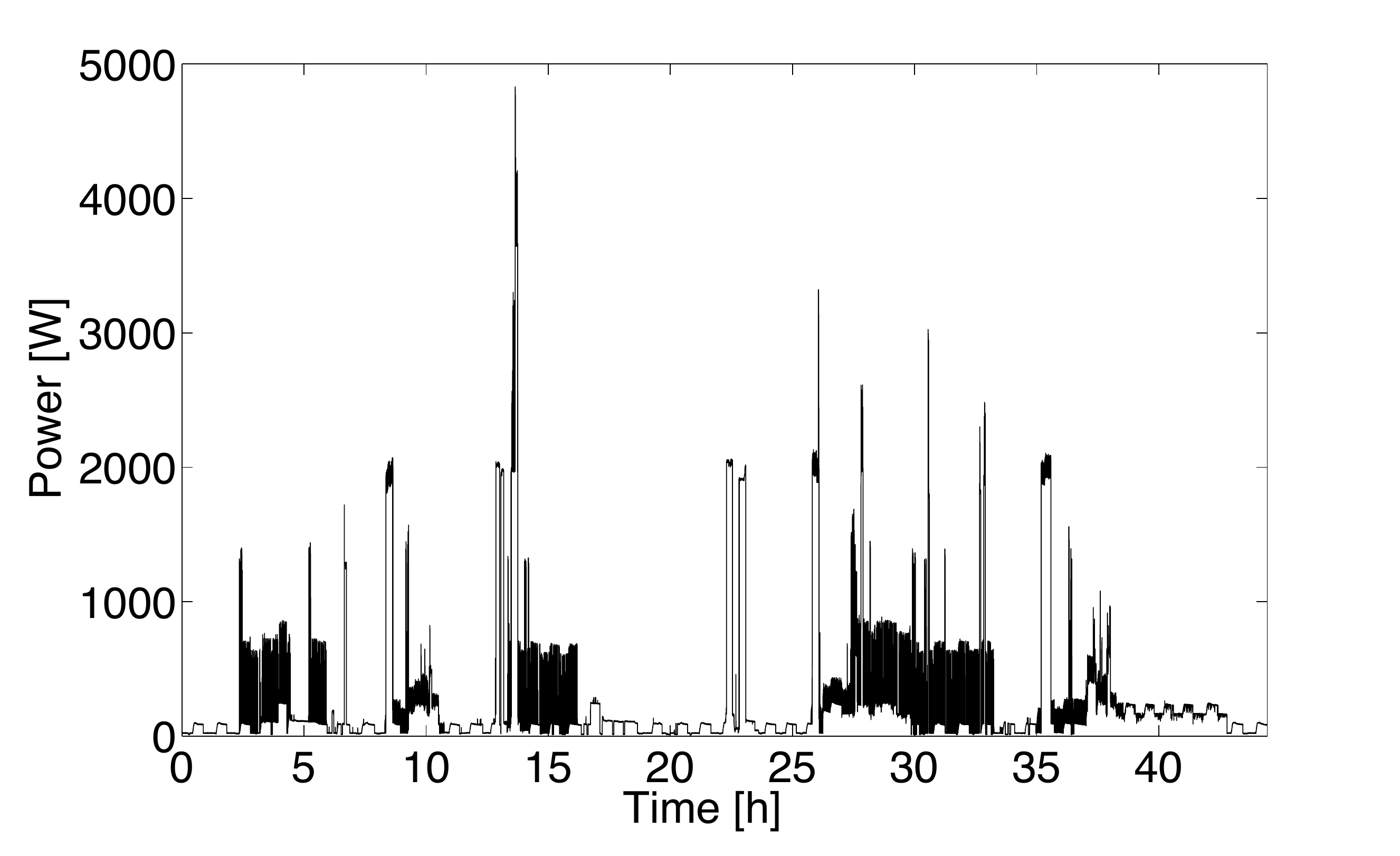}
	\caption{Time section of original household's power draw}
		\label{fig:originalTS}
\end{minipage}
\begin{minipage}{0,32\columnwidth}
		\includegraphics[scale=0.25]{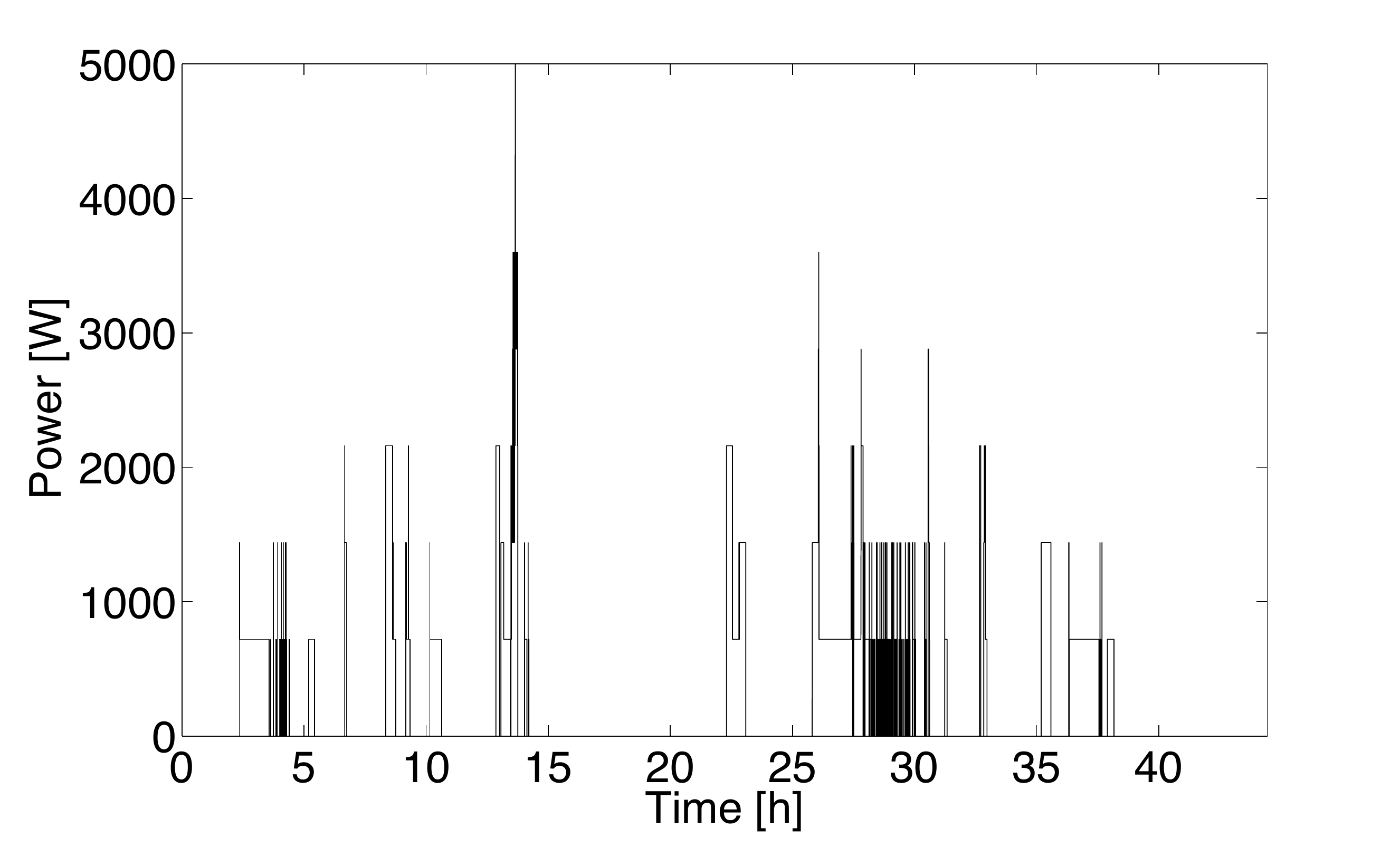}
	\caption{Time section of original household's power draw obfuscated by \ac{BLH}}
		\label{fig:blhTS}
\end{minipage}
\begin{minipage}{0,32\columnwidth}
		\includegraphics[scale=0.25]{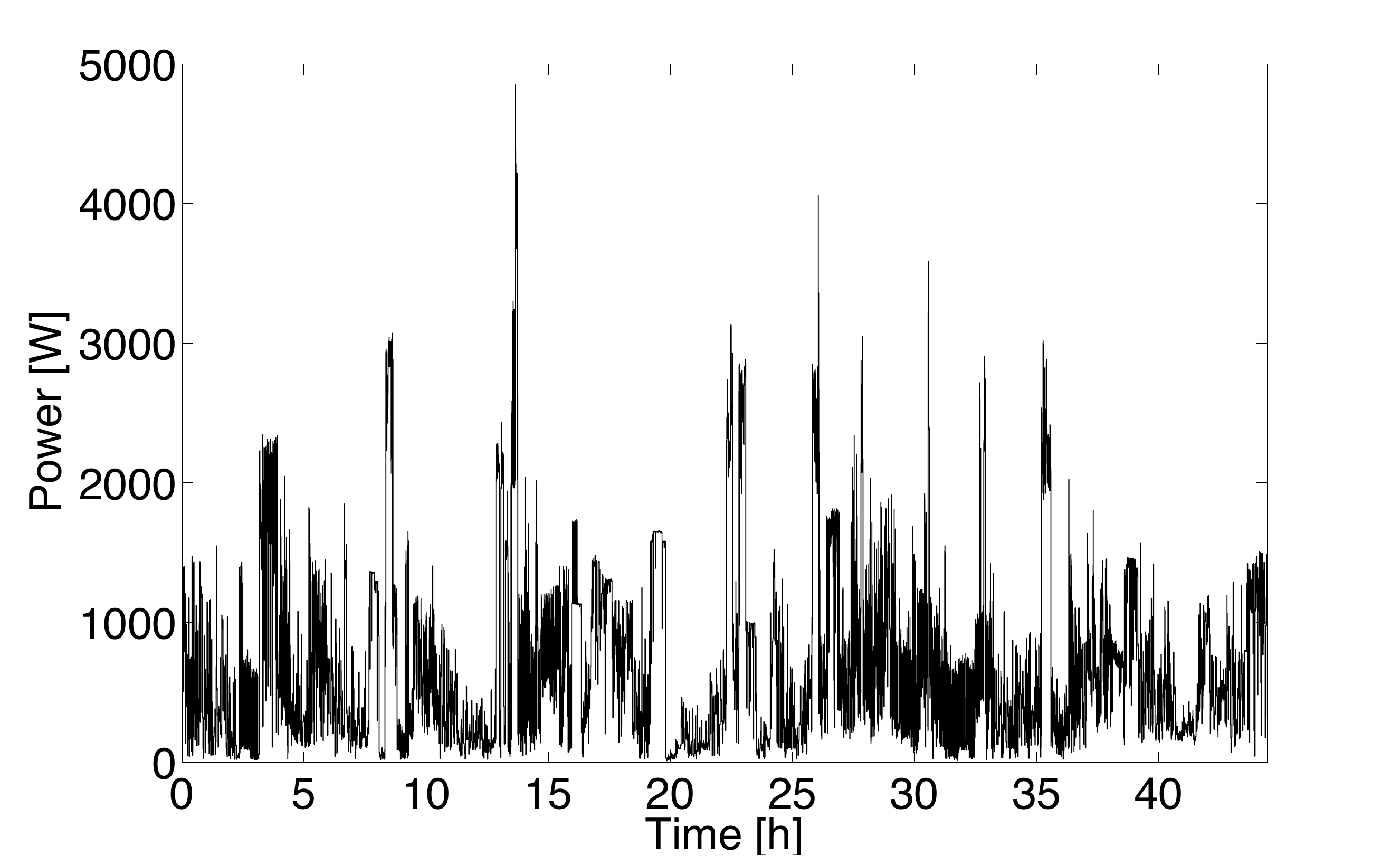}
	\caption{Time section of original household's power draw obfuscated by \ac{LLH}}
		\label{fig:llhTS}
\end{minipage}
\end{figure*}
\end{landscape}

\subsection{\ac{NILM} on obscured power draw}
To evaluate the ability of \ac{BLH} and \ac{LLH} to obfuscate the household draw to the extent that NILM algorithms do not work any more, we tested the obfuscated demand profile with the presented \ac{NILM} algorithm.
In the case of \ac{BLH} we used the same battery capacities as before $C \in [10,70,100,200,400,600] \: Ah$.
In Table \ref{tab:BLHResult} the \ac{ACC} results of the load disaggregator on appliance level and in total are presented. 
As reference, the load disaggregator result for the non-obfuscated case and the
result of a load disaggregator estimating all appliances to be off over the whole observation window are listed.
As expected, the \ac{ACC} decreases with an increased battery capacity. 
The results show that different appliances are affected in a different way. 
For example the TV or the fridge having a comparable low energy demand, are highly affected by the \ac{BLH} algorithm in which energy hogs such as the coffee machine or the hoover are much harder to hide.
By comparing the \ac{ACC} results with the reference \ac{ACC} results, \ac{BLH} can obfuscate the total power demand very well.
A battery of size $100\:Ah$ is sufficient to make the load disaggregator not working in a sufficient way any longer.
\begin{landscape}
\begin{table*}
\centering
\begin{tabular}{|c|ccccccc|c|} 
\hline
&\multicolumn{7}{c|}{\textbf{\ac{ACC} on appliance level}} & \\
\hline

  & TV & coffee machine  & dishwasher & fridge & hoover & water kettle & washing machine & \\
  \hline
states  & 3 & 2  & 2 & 4 & 2 & 2 & 4 & \\ 
power [W] & [0 10 160] & [0 1280]  & [0 1900] & [0 8 80 230]  & [0 1200]  & [0 1700]   & [0 130 240 1920]  & \\
\hline
\hline
\multicolumn{7}{c}{}\\
\hline
\textbf{C} & \multicolumn{7}{c|}{\textbf{\ac{BLH} case}} &  \textbf{\ac{ACC} total} \\
\hline
10Ah & 0,56 & 	0,99 & 	0,95 & 	0,69 & 	0,99 & 	0,98 &	0,79 & 0,84 \\
70Ah &  0,41& 	0,99 & 	0,95 & 	0,55 & 	0,99 & 	0,98 & 	0,89 & 	0,82 \\
100Ah & 0,05 & 	0,98 & 	0,95 & 	0,5 & 	0,99 &	0,98 & 	0,71 & 0,74 	\\
200Ah &  0,19 & 	0,97 & 	0,95 &	0,46 & 	0,99 & 	0,99 & 	0,78 & 0,76	\\
400Ah & 0,12 & 	0,89 & 	0,95 & 	0,42 & 	0,94 & 	0,99 &	0,88 	& 0,74\\
600Ah & 0,07 & 	0,98 & 	0,95 & 	0,42 & 	0,99 & 	0,97 & 	0,87 & 0,75\\
\hline

\multicolumn{7}{c}{}\\
\hline
\textbf{daily consumption} & \multicolumn{7}{c|}{\textbf{\ac{LLH} case}} &  \textbf{\ac{ACC} total} \\
\hline
2.5kWh & 0,62	& 0,98& 	0,93 &	0,52  & 	0,98 & 	0,98 &	0,60 & 0,8 \\
5kWh &  0,61& 	0,97& 	0,95 & 	0,51 & 	0,95 & 	0,98 & 	0,51 & 0,78  	\\
7.5kWh & 0,60 & 	0,95& 	0,95 & 	0,50 & 	0,92 & 	0,98 & 	0,44 & 0,76 	\\
10kWh &  0,60 & 	0,92 & 	0,95& 	0,51 &	0,88 & 	0,98 & 	0,43 & 0.75 	\\
\hline

\multicolumn{7}{c}{}\\
\hline
\textbf{reference} & 0,60 &	0,92 & 	0,95 &	0,52 &	0,88 &	0,98 &	0,43 & 0,75 \\
 \hline
\textbf{original} &  0,68 &   0,98 &    0,95 &   0,88 &    0,99 &    0,99 &    0,89 & 0,91 \\ 
 \hline
\end{tabular}
\caption{Accuracy of proposed \ac{NILM} algorithm for \ac{BLH} and  \ac{LLH} disguise household power draw on appliance level and on total. As a comparison, the \ac{ACC} results on the original power draw and a reference estimation, where all appliance states are off, are listed as well.}
\label{tab:BLHResult}
\end{table*}
\end{landscape}
In the case of an \ac{LLH} obfuscated power demand, the power time series is inferred by a modified beta distribution controlled boiler noise signal. 
Table \ref{tab:BLHResult} lists the \ac{ACC} results on appliance level and in total.
By changing the energy target consumption, the \ac{ACC} decreases whereas the privacy increases.
In case $4$, we even get the same \ac{ACC} as in the reference case assuming all devices to be off over the whole observation window.


\section{Discussion}\label{sec:discussion}

To use \ac{BLH} and \ac{LLH} in real households, additional hardware and devices are needed.
In case of \ac{BLH} an adequate battery, additional wiring and a controlling unit including an inverter has to be installed. 
Assuming the implementation using a cheap maintenance free starter battery of a car with a short life cycle, the approximated costs for this battery are approximately 150\euro\: for a 100\:Ah type, that must be renewed after a few years. A fully remotely controllable inverter/charger combination for currents up to 35\:A costs approximately 1000\euro\:\cite{Prokop2014}. Compared to the battery system with the inverter/charger combination the measuring units and control system are cheap, especially when considering mass production. Hence, the initial total costs of a 100\:Ah \ac{BLH} system should be in the range of 1000\euro\: to 1500\euro\:\cite{Prokop2014}.
Compared to \ac{BLH}, \ac{LLH} has a lower need for additional hardware.
No additional battery and wiring is needed, in which the boiler requires a controlling unit to adjust the power values of the boiler.
For a \ac{LLH} the approximated costs are less than 200\euro \: assuming an already installed electric hot water boiler. A power regulator for 230\:V and a maximum power of 2000\:VA is available for less than 50\euro, similarly to \ac{BLH} the control system should be quite cheap, but the tricky part of a \ac{LLH} implementation is the installation of a temperature sensor in an existing hot water boiler, which could be more expensive.
According the work of Monacchi et. al. \cite{monacchi2013Nov} there are regional differences of appliance type usage.
In Italy, electric boilers are quite uncommon whereas electric boilers in Austria are very usual. 
Thus, dependent on the region different load hiding techniques can be applied (Austria-\ac{LLH}, Italy-\ac{BLH}). 
Both, \ac{BLH} and \ac{LLH} are trying to obfuscate the household's power demand as much as possible. 
Examples for the original power draw (Figure~\ref{fig:originalTS}), the \ac{BLH} (Figure~\ref{fig:blhTS}) and the \ac{LLH} (Figure~\ref{fig:llhTS}) obfuscated consumption data are presented.
Taking the results of the previous section into consideration, \ac{BLH} is better than \ac{LLH} in obfuscating the total power demand. The \ac{RMSE} and the \ac{ACC} values of a \ac{BLH} system are either the same or better than for a \ac{LLH} system.
But \ac{BLH} has the disadvantage that it is not able to obfuscate energy-intensive appliances without installing very costly batteries.
In contrast, \ac{LLH} is modifying the power demand in a way that the presented \ac{NILM} cannot detect running states of both small consuming devices as well as of energy hogs. 
The results of Table \ref{tab:BLHResult} show that the estimator get nearly the same results than a load disaggregator estimation all appliances to be off. 
Therefore, \ac{LLH} is promising as a load hiding technique due to its ability to obfuscate appliance and household consumption data.
\section{Conclusion}\label{sec:conclusion}
To ensure privacy in homes is an important topic for households with smart meters.
Smart meters are providing fine-grained consumption data in which it is possible to extract habits and behaviors of inhabitants by assigning a meaning to the consumption data.
The technique called \acf{NILM} is the basic step to get appliance usage data from the household power consumption which is finally used for user modeling. 
This information introduces privacy threats which are tried to be preserved by so-called load hiding techniques.
State-of-the-art load hiding techniques are obfuscating the consumption data by using a battery to modify the to the energy meter visible energy by adding or withdrawing energy.
The technique is called \acf{BLH}. 
In this paper, a novel load hiding technique based on a controllable household device (e.g., electric boiler) is presented. 
The used household device (such as the electric boiler) should be non-user-driven and should have a daily consumption demand.
It tries to obfuscate the power demand by randomly affecting the household's power demand using noise, whereas \ac{BLH} tries to flatten the power demand as much as possible.
In this paper, both load hiding techniques are compared by the \ac{RMSE} value of the obfuscated and non-obfuscated power consumption and by the applicability of \ac{NILM} algorithms on the obfuscated power demand.
Simulation results on real household data show that both techniques strengthen the household privacy in a way that the used \ac{NILM} approach is disabled to identify running appliances.
Although for a given energy turnover the presented \ac{BLH} system achieves a better behavior for a household with mostly small appliances, the \ac{LLH} is better in obfuscating appliances with high power consumption.
If a suitable device, e.g. an electric boiler, is available, a suitable \ac{LLH} system can be installed of much lower cost.
\bibliographystyle{IEEEtran}
\bibliography{dominik}

\end{document}